\documentclass[pra,twocolumn,superscriptaddress,showpacs,floatfix]{revtex4-1}

\usepackage[latin1]{inputenc}
\usepackage{exscale}
\usepackage[T1]{fontenc}
\usepackage{epsfig}
\usepackage{graphicx}
\usepackage{amsmath}
\usepackage{enumerate}

\newcommand{\ee}{\mathrm{e}}
\newcommand{\ii}{\mathrm{i}}
\newcommand{\dd}{\mathrm{d}}

\newcommand{\jj}{\mathcal{J}}

\newcommand{\an}[1]{\hat{#1}}
\newcommand{\cre}[1]{\hat{#1}^\dag}

\renewcommand{\baselinestretch}{0.98}

\begin{document}

\title{Phonon resonances in atomic currents through Bose-Fermi mixtures in optical lattices}

\author{M. Bruderer}
\affiliation{Fachbereich Physik, Universität Konstanz, D-78457 Konstanz, Germany}

\author{T. H. Johnson}
\affiliation{Clarendon Laboratory, University of Oxford, Parks Road, Oxford OX1 3PU,
United Kingdom}

\author{S. R. Clark}
\affiliation{Centre for Quantum Technologies, National University of Singapore, 3 Science
Drive 2, Singapore 117543}
\affiliation{Clarendon Laboratory, University of Oxford, Parks Road, Oxford OX1 3PU,
United Kingdom}

\author{D. Jaksch}
\affiliation{Clarendon Laboratory, University of Oxford, Parks Road, Oxford OX1 3PU,
United Kingdom}
\affiliation{Centre for Quantum Technologies, National University of Singapore, 3 Science
Drive 2, Singapore 117543}

\author{A. Posazhennikova}
\affiliation{Fachbereich Physik, Universität Konstanz, D-78457 Konstanz, Germany}

\author{W. Belzig}
\affiliation{Fachbereich Physik, Universität Konstanz, D-78457 Konstanz, Germany}

\date{\today}

\begin{abstract}
We present an analysis of Bose-Fermi mixtures in optical lattices for the
case where the lattice potential of the fermions is tilted and the bosons
(in the superfluid phase) are described by Bogoliubov phonons. It is shown
that the Bogoliubov phonons enable hopping transitions between fermionic
Wannier-Stark states; these transitions are accompanied by energy
dissipation into the superfluid and result in a net atomic current along the
lattice. We derive a general expression for the drift velocity of the
fermions and find that the dependence of the atomic current on the lattice
tilt exhibits negative differential conductance and phonon resonances.
Numerical simulations of the full dynamics of the system based on the
time-evolving block decimation algorithm reveal that the phonon
resonances should be observable under the conditions of a realistic
measuring procedure.
\end{abstract}



\maketitle


\section{Introduction}

One of the most intriguing prospects opened up by recent advances in atomic
physics is the possibility of studying many-body quantum systems. In
particular, ultracold atoms confined to optical lattice potentials have been
shown to be perfectly suitable for implementing physical models of
fundamental interest not only to the field of atomic physics but also to
condensed matter physics~\cite{Lewenstein-AiP-2006,Bloch-2007}. Specific
examples of a highly versatile many-body system include Bose-Fermi mixtures
in optical lattices, which have been used recently to analyze the effect of
fermionic impurities on the superfluid to Mott-insulator
transition~\cite{Guenter-PRL-2006,Ospelkaus-PRL-2006,Best-PRL-2009}. A
further experimental setup closely related to condensed matter systems
consists of ultracold atoms in \emph{tilted} optical lattice potentials.
Several fundamental quantum mechanical processes related to nonequilibrium
transport of particles have been observed in this setup such as, e.g.,
Landau-Zener tunneling~\cite{Zenesini-AX-2007}, Bloch
oscillations~\cite{Dahan-PRL-1996,Roati-PRL-2004}, and processes analogous
to photon-assisted tunneling~\cite{Sias-AX-2007}.

Of particular importance, collisionally induced transport of fermions
confined to an optical lattice and coupled to an ultracold bosonic bath has
been observed in an experimental setup of a similar type to the one
considered in this article~\cite{Ott-PRL-2004}. Furthermore, a closely
related theoretical analysis of collision-induced atomic currents along a
tilted optical lattice was provided by
Ponomarev~\emph{et~al.}~\cite{Ponomarev-PRL-2006} based on a random matrix
approach. In essence it was pointed out in Ref.~\cite{Ponomarev-PRL-2006}
that the atomic current exhibits Ohmic and negative differential conductance
(NDC).

In the present article we study a natural extension of the mentioned
experiments, namely lattice Bose-Fermi mixtures with the fermions confined
to a tilted optical potential. Motivated by earlier
considerations~\cite{Bruderer-PRA-2007,Bruderer-NJP-2008} we show that this
system is highly appropriate for exploring the effects of electron-phonon
interactions on nonequilibrium electric transport through a conductor.
Specifically, we demonstrate that electron-phonon resonances, predicted to
exist in solids nearly 40~years
ago~\cite{Bryxin-SSC-1972,Dohler-SSC-1975,Bryksin-JPCM-1997}, yet for which
there seems to be no conclusive experimental
evidence~\cite{Feng-PRB-2003,Leo-2003}, can be realized using ultracold
atoms.

\begin{figure}[h!]
\begin{center}
  \includegraphics[height = 115pt]{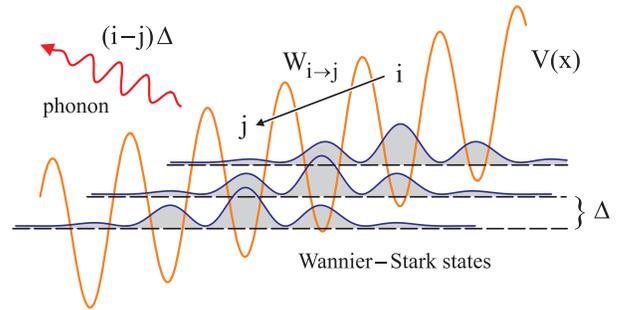}
  \caption{(Color online) Fermions confined to a tilted potential $V(x)$ form a Wannier-Stark ladder
    with a constant energy separation~$\Delta$ between adjacent lattice sites.
    The fermion-boson interaction enables phonon-assisted transitions from site $i$ to $j$
    at the rate~$W_{i\rightarrow j}$. The released energy $(i-j)\Delta$ is dissipated
    in the form of a phonon emitted into the superfluid.}\label{scheme}
\end{center}
\end{figure}\vspace{5pt}

In our model the fermions take the role of electrons, whereas the bosons
provide a nearly perfect counterpart to acoustic phonons in solids. The tilt
imposed on the lattice potential of the fermions corresponds to an applied
bias voltage to the system.  The collisions between fermions and bosons
result in fermion-phonon relaxation processes, which give rise to a net
atomic current along the tilted lattice
potential~\cite{Ponomarev-PRL-2006,Bruderer-NJP-2008,Kolovsky-PRA-2008}, as
illustrated in Fig.~\ref{scheme}. We show that in our case, where both the
fermions and bosons are trapped in an optical lattice, phonon resonances in
the atomic current occur. They arise as the momentum of the phonon emitted
in the relaxation process approaches a so-called Van Hove
singularity~\cite{VanHove-PR-1953} at the upper edge of the phonon band.
Moreover, we shall see that phonon resonances, at least on the level of
approximations made in this article, are also expected to occur in Bose-Bose
mixtures~\cite{Catani-PRA-2008}.

There are crucial advantages of our cold atom implementation over a
solid-state system: First, neither impurities nor imperfections in the
system suppress the resonances in the current~\cite{Feng-PRB-2003} and hence
fermion-phonon scattering is the only relaxation process, which can be fully
controlled via the bosonic system
parameters~\cite{Best-PRL-2009,Ernst-NP-2009}. Second, large lattice tilts
can be achieved with an energy mismatch between neighboring fermion sites
that exceeds the bandwidth of both the fermions and the phonons. This makes
it possible to study the influence of the phonon density of states on the
atomic current over the entire phonon band. In addition, it allows us to
observe negative differential
conductance~\cite{Ponomarev-PRL-2006,Bruderer-NJP-2008,Kolovsky-PRA-2008},
which is realized in a solid-state system with difficulty by resorting to
semiconductor superlattices~\cite{Esaki-IBM-1970,Dohler-SSC-1975,Leo-2003}.
Last, the parameters of our system can be chosen to ensure that Landau-Zener
tunneling to higher fermion bands---often a significant effect in high field
transport~\cite{Leo-2003}---is negligible despite the large lattice
tilts~\cite{Bruderer-NJP-2008} justifying the use of a tight-binding
framework.

The key experimental techniques required for our scheme are twofold: first,
the independent trapping of atoms of different species in species-specific
optical lattice potentials~\cite{Mandel-PRL-2003,Lamporesi-PRL-2010} and,
second, the tunability of the boson-boson and boson-fermion interactions
by Feshbach resonances~\cite{Roati-PRL-2007,Best-PRL-2009}. Also, we exploit
the possibility of implementing low-dimensional systems by strongly
increasing the depth of the optical potentials along specific directions. In
this way effectively one-dimensional systems can be realized by tightly
confining atoms to an array of tubes~\cite{Bloch-2007}.

The structure of this article is as follows: In Sec.~II we start from the
Bose-Fermi Hubbard model and outline the effective description of the
Bose-Fermi mixture in terms of Bogoliubov phonons and Wannier-Stark states.
In Sec.~III we derive a general expression for the drift velocity of the
fermions and show that this expression encompasses the phenomena of negative
differential conductance and phonon resonances. Section~IV contains the
results of a near-exact numerical simulation of a realistic experimental
procedure to determine the dependence of the atomic current on the lattice
tilt. We end with the conclusions in Sec.~V.


\section{Effective model}

The specific system we consider consists of a homogeneous, one-dimensional
Bose-Fermi mixture of bosons and spin-polarized fermions, both trapped in
\emph{separate} optical lattice potentials. If the potentials are
sufficiently deep so that only the lowest Bloch band is occupied then the
Bose-Fermi mixture can be described by the Bose-Fermi Hubbard model~\cite{Albus-PRA-2003}
\begin{equation}\label{ham}
\begin{split}
    \hat{H}_{bf} = & - J_b\sum_{<i,j>}\cre{a}_i\an{a}_j - J_f\sum_{<i,j>}\cre{c}_i\an{c}_j + \sum_j(j\Delta)\cre{c}_j\an{c}_j\\
    & + \frac{1}{2}U_{b}\sum_j \cre{a}_j\cre{a}_j\an{a}_j\an{a}_j + U_{bf}\sum_j\cre{a}_j\an{a}_j\cre{c}_j\an{c}_j\,,
\end{split}
\end{equation}
where $\langle i,j\rangle$ denotes the sum over nearest neighbors. The
operators $\cre{a}_j$ ($\an{a}_j$) create (annihilate) a boson and,
similarly, the operators $\cre{c}_j$ ($\an{c}_j$) create (annihilate) a
spinless fermion in a Wannier state localized at site $j$. The bosonic and
fermionic hopping parameters are $J_b$ and $J_f$, respectively, and the
on-site boson-boson and boson-fermion interactions are characterized by
the energies $U_{b}$ and $U_{bf}$, both positive and independently tunable.
In contrast to the bosons, the lattice potential of the fermions is assumed
to be tilted with an energy splitting~$\Delta\geq 0$ between adjacent sites.

\begin{figure}[t]
\begin{center}
 \includegraphics[height = 100pt]{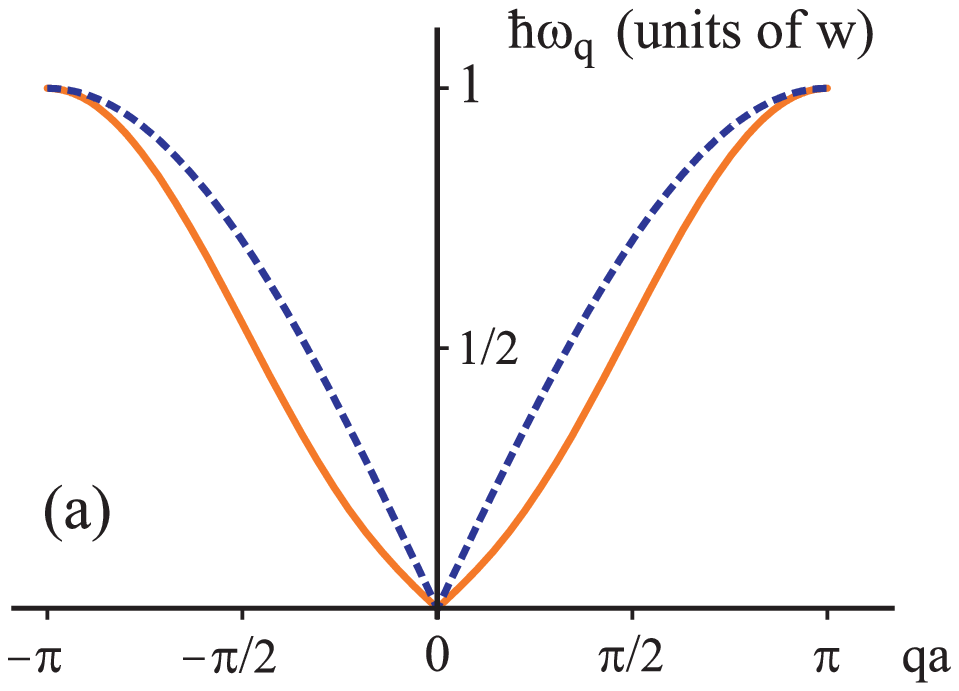}\hspace{3pt}\includegraphics[height = 100pt]{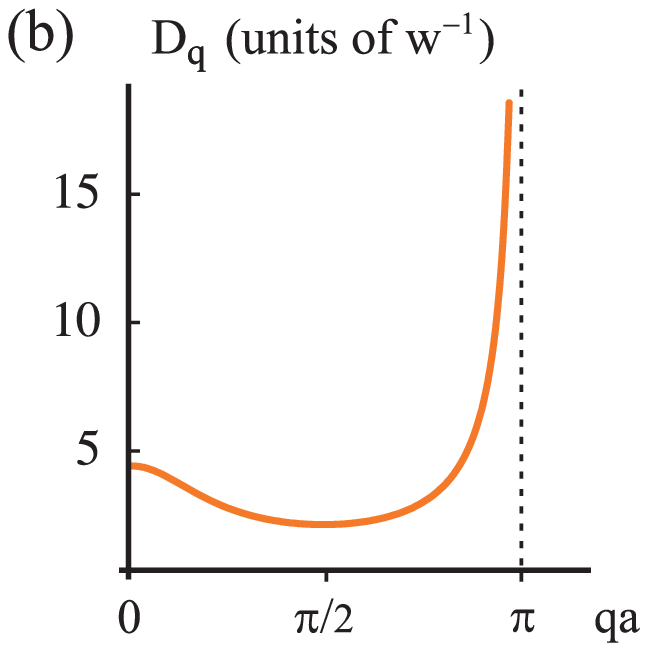}
  \caption{(Color online) (a)~The dispersion relation of acoustic phonons in a solid with
   $\hbar\omega_q\sim|\sin(qa/2)|$ (dashed line) and of
   Bogoliubov phonons in a superfluid confined to an optical lattice
   (full line) in units of the bandwidth $w = 4J_b\sqrt{1+U_bn_b/2J_b}$ for the parameters $U_b n_b/J_b = 0.5$.
   (b)~The corresponding density of states $D_q = |\partial\hbar\omega_q/\partial(qa)|^{-1}$ (full line)
   of the Bogoliubov phonons is approximately constant for small momenta and
   exhibits a Van Hove singularity near the edge of the first Brillouin zone (dashed line).}\label{phonon}
\end{center}
\end{figure}

At low temperatures and for sufficiently small boson-boson interactions
$U_{b}/J_b\sim 1$ most bosons are in the superfluid phase and accurately
described by the phononic excitations of the superfluid. This description is
obtained by transferring the bosonic part of the Hamiltonian~$\hat{H}_{bf}$
into momentum space and adopting the Bogoliubov
approach~\cite{Oosten-PRA-2001}, which results in the phonon Hamiltonian
\begin{equation}
  \hat{H}_{ph} = \sum_q\hbar\omega_q\cre{b}_q\an{b}_q\,.
\end{equation}
Here, $q$ is the momentum running over the first Brillouin zone,
i.e.,~$-\pi/a<q\leq\pi/a$ with $a$ the lattice spacing, and the bosonic
operators $\cre{b}_q$ ($\an{b}_q$) create (annihilate) a Bogoliubov phonon.
The excitation spectrum of the phonons is given by the Bogoliubov dispersion
relation $\hbar\omega_q = \sqrt{E_q(E_q + 2U_b n_b)}$, where $n_b$ is the
bosonic occupation number and $E_q = 4J_b\sin^2(qa/2)$ is the dispersion
relation of noninteracting bosons in the optical lattice. The dispersion of
the Bogoliubov phonons, albeit not identical, is remarkably similar to the
dispersion of acoustic phonons in a solid-state system with
$\hbar\omega_q\sim|\sin(qa/2)|$, shown in Fig.~\ref{phonon}(a). The common
features are a linear dispersion in the limit $q\rightarrow 0$ and the band
structure with a gap at the boundary of the first Brillouin zone. It should
be pointed out that the accuracy of the Bogoliubov description of phonons
has been demonstrated experimentally~\cite{Ernst-NP-2009}, thus we have
perfect knowledge of and control over the phonons in our system.

The fermions confined to the tilted potential are represented by the
eigenstates of the fermionic part of $\hat{H}_{bf}$, i.e.,~the Wannier-Stark
states~\cite{Emin-PRB-1987,Gluck-PR-2002}. The unitary transformation that
relates the Wannier operators~$\an{c}_j$ to the Wannier-Stark
operators~$\an{d}_j$ is given by
\begin{equation}\label{bestrans}
    \an{d}_j = \sum_i\jj_{i-j}(2J_f/\Delta)\an{c}_i\,,
\end{equation}
where $\jj_n(z)$ are Bessel functions of the first kind~\cite{Watson-2008}.
It follows immediately from the properties of the Bessel functions
$\jj_n(z)$ that the Wannier-Stark states are centered at lattice site $j$
and have a width of the order of $J_f/\Delta$. By invoking the identities
for the Bessel functions $\sum_n\jj_{n-m}(z)\jj_{n-m^\prime}(z) =
\delta_{m,m^\prime}$ and $(2n/z)\,\jj_n(z) = \jj_{n+1}(z) + \jj_{n-1}(z)$
one finds that the transformed fermionic part of $\hat{H}_{bf}$ takes the
diagonal form
\begin{equation}
  \hat{H}_{ws} = \sum_j(j\Delta)\cre{d}_j\an{d}_j\,.
\end{equation}
The spectrum of $\hat{H}_{ws}$ constitutes a so-called Wannier-Stark
ladder~\cite{Emin-PRB-1987,Gluck-PR-2002} with a constant energy
separation~$\Delta$ between adjacent sites, shown in Fig.~\ref{scheme}. We
emphasize that the Wannier-Stark states are stationary and hence a net
current of fermions along the tilted lattice only develops in presence of
additional relaxation processes. In particular, it results from the inverse
of the transformation in Eq.~(\ref{bestrans}) that a fermion initially
localized at a single lattice site undergoes coherent Bloch oscillations
with frequency~$\Delta/\hbar$~\cite{Dahan-PRL-1996}.

To rewrite the boson-fermion interaction in terms of Bogoliubov phonons and
Wannier-Stark states we apply the same transformations as for the purely
bosonic and fermionic parts of $\hat{H}_{bf}$. Invoking the results
established in Refs.~\cite{Oosten-PRA-2001,Emin-PRB-1987} we find the
fermion-phonon interaction Hamiltonian
\begin{equation}\label{hint}
    \hat{H}_{int} = U_{bf}n_b\sum_j\cre{d}_j\an{d}_j
    + \sum_{j,\ell,q}\left[f_{q,\ell}\an{b}_q\ee^{-\ii qaj} + \mathrm{h.c.}\right]
    \cre{d}_{j+\ell}\an{d}_j
\end{equation}
with the corresponding matrix elements
\begin{equation}\label{matrix}
    f_{q,\ell} =  U_{bf}\left(\frac{n_b E_q}{N\hbar\omega_q}\right)^{1/2}
    \jj_\ell\left(\frac{4J_f}{\Delta}\sin\frac{qa}{2}\right)\ii^\ell\ee^{-\ii qa\ell/2}\,,
\end{equation}
where $N$ is the number of lattice sites. We note that $\hat{H}_{int}$
coincides exactly with the description of the electron-phonon interactions
in a solid-state system~\cite{Emin-PRB-1987} and thus is perfectly adequate
for investigating related phonon effects.

The fermion-phonon interaction $\hat{H}_{int}$ describes the creation of a
Bogoliubov phonon out of the superfluid phase and the reverse process, both
caused by the hopping process of the fermions. These incoherent processes
involving a single phonon enable fermions to make transitions between
Wannier-Stark states separated by $\ell$ lattice sites. On the other hand,
coherent processes resulting from $\hat{H}_{int}$, in which a virtual phonon
is emitted and reabsorbed, lead to a phonon-mediated fermion-fermion
interaction, a renormalized fermion hopping and a mean-field energy shift of
the fermions~\cite{Bruderer-PRA-2007,Bruderer-NJP-2008}.

The effective fermion-fermion interaction, being short range, can be safely
neglected if we assume a filling factor of the fermions much lower than~1.
However, we have to take into account the renormalization of the fermion
hopping~$J_f$ when comparing our theoretical model to experimental or
numerical results. As expected from theoretical
considerations~\cite{Bruderer-PRA-2007,Bruderer-NJP-2008} and confirmed by
the numerical results in Sec.~IV, the renormalization reduces the bare
hopping. Last, the mean-field energy shift is explicitly given by $U_{bf}n_b
+ \sum_q|f_{q,0}|^2/\hbar\omega_q$, which is readily determined in the
regime $U_bn_b/J_b\ll 1$ by taking the continuum limit~\footnote{Explicitly,
the continuum limit of the sum over all momenta in the lowest band is
$(Na)^{-1}\sum_q\rightarrow \int_{-\pi/a}^{\pi/a}\dd q/2\pi$.}. For a single
delocalized fermion with $J_f/\Delta\gg 1$ the shift is simply $U_{bf}n_b$,
whereas for a localized fermion with $J_f/\Delta\ll 1$ we find
$U_{bf}n_b(1-a U_{bf}/2\hbar c_s)$, with $c_s = (a/\hbar)\sqrt{2J_bU_bn_b}$
the speed of sound in the superfluid.

It is essential for our model that neither interactions with the fermions
nor the trapping potential confining the fermions destroy the phononic
excitations of the superfluid. These conditions are met by using
species-specific optical lattice
potentials~\cite{Mandel-PRL-2003,Lamporesi-PRL-2010} and by limiting the
number of fermions in the
system~\cite{Guenter-PRL-2006,Ospelkaus-PRL-2006,Best-PRL-2009}. Moreover,
we restrict our analysis to fermions moving slower than superfluid critical
velocity, which is $c_s$ according to Landau's
criterion~\cite{Landau-JPU-1941}, to avoid excitations other than those
caused by the hopping transitions. More precisely we consider the parameter
regime $\zeta = 2aJ_f/\hbar c_s\ll 1$, where $2aJ_f/\hbar$ is the maximal
group velocity of the fermions in the lowest Bloch band.


\section{Atomic currents}

Since the number of fermions is significantly smaller than number of bosons
we effectively treat the superfluid as a phonon bath. Accordingly we
describe the dynamics of the fermions by a master equation for the
probabilities $P_j$ that a fermion occupies a Wannier-Stark state at
site~$j$. The master equation reads
\begin{equation}\label{master}
    \partial_t P_i = \sum_j\left[P_j(1-P_i)W_{j\rightarrow i} - P_i(1-P_j)W_{i\rightarrow j}\right]\,,
\end{equation}
where $W_{i\rightarrow j}$ are the rates for a phonon-assisted transition
from site $i$ to site $j$ and the factors~$(1 - P_j)$ take the Pauli
exclusion principle into account. The probabilities~$P_i$ either describe
the occupation of a single fermion or the distribution of an ensemble of
fermions; in both cases we choose $\sum_j P_j = 1$. The average position
$\bar{x}$ of the fermions at time $t$ is accordingly $\bar{x} =
\sum_j(aj)P_j$ and the average drift velocity $\bar{v}_d$, which quantifies
the atomic current along the lattice, is given by $\bar{v}_d =
\partial_t\bar{x} = \sum_j(aj)\partial_t P_j$.

To obtain a useful expression for the drift velocity $\bar{v}_d$ we exploit
the fact that the system is homogeneous so that the transition rates
$W_{i\rightarrow j}$ only depend on the jump distance $\ell = i-j$, where
jumps with $\ell > 0$ are defined to be downward the tilted lattice, as
depicted in Fig.~\ref{scheme}. By using Fermi's golden rule based on the
interaction Hamiltonian $\hat{H}_{int}$, i.e.,~to second order in the
coupling $U_{bf}$, we find
\begin{equation}\label{wsum}
\begin{split}
    W^E_{\ell} &= \frac{2\pi}{\hbar}\sum_q|f_{q,\ell}|^2 (N_q + 1)\,\delta\left(\hbar\omega_q - \ell\Delta\right)\,,\\
    W^A_{\ell} &= \frac{2\pi}{\hbar}\sum_q|f_{q,\ell}|^2 N_q\,\delta\left(\hbar\omega_q - \ell\Delta\right)\,,
\end{split}
\end{equation}
where $N_q = (\ee^{\hbar\omega_q/k_B T}-1)^{-1}$ is the mean number of
phonons with momentum $q$ in the superfluid at temperature~$T$ and $k_B$ is
Boltzmann's constant. The rate $W^E_{\ell}$~($W^A_{\ell}$) corresponds to
the process, where a fermion jumps $\ell$~sites down (up) the lattice and
thereby emits (absorbs) a single phonon of energy $\ell\Delta$. The
expressions for the jump rates $W^E_{\ell}$ and $W^A_{\ell}$ are valid as
long as heating effects caused by the emitted phonons are negligible.

To determine $\bar{v}_d$ in terms of transition rates we substitute the
expression in Eq.~(\ref{master}) for $\partial_t P_j$ and find after some
rearrangement that
\begin{equation}\label{master2}
    \bar{v}_d = \sum_{j,\ell>0} (a\ell) \left[W^0_\ell P_j(1-P_{j+\ell}) - W^A_\ell P_j(P_{j+l} -P_{j-\ell})\right]
\end{equation}
with $W^0_\ell = (2\pi/\hbar)\sum_q|f_{q,\ell}|^2\,\delta\left(\hbar\omega_q
- \ell\Delta\right)$ the phonon emission rate at zero temperature. The first
term in the sum in Eq.~(\ref{master2}) vanishes if the fermions are
degenerate at zero temperature with all lattice sites $j \geq j_c$ occupied,
i.e.,~$P_j = 1$ for $j\geq j_c$ and $P_j = 0$ for $j<j_c$ with $j_c$ fixed
by the number of fermions, whereas the second term represents finite
temperature effects. For the system far from degeneracy and for small
fermion occupation numbers we may make the approximation $P_i P_j \approx 0$
and use the condition $\sum_j P_j = 1$, which yields $\bar{v}_d =
\sum_{\ell>0}(a\ell)\,W^0_\ell$ for the drift velocity. We note that at this
level of approximation the expression for the drift velocity and all
subsequent analytical considerations apply, mutatis mutandis, also to
lattice Bose-Bose mixtures~\cite{Catani-PRA-2008}.

Taking the continuum limit of the sum over all phonon momenta (appearing in
$W^0_\ell$) and integrating over the first Brillouin zone we finally obtain
\begin{equation}\label{vdres}
    \bar{v}_d = \frac{a^2N}{\hbar}\sum_{\ell>0}^{\ell_{\mathrm{max}}}\,\ell\,\left|\frac{\partial\hbar\omega_q}{\partial q}\right|^{-1}
    |f_{q,\ell}|^2\:\bigg|_{q=q_\ell}\:,
\end{equation}
with $q_\ell$ implicitly defined by $E_{q_\ell} = \sqrt{(\ell\Delta)^2 +
(U_bn_b)^2} - U_bn_b$. The latter relation also yields the maximum jump
distance~$\ell_{\mathrm{max}}$, which is compatible with energy conservation
in the fermion-phonon relaxation process. We see from Eq.~(\ref{vdres})
that the total drift velocity $\bar{v}_d$ is determined by the sum over all
admissible jump distances~$\ell$ weighted by the phonon density of states
$|\partial\hbar\omega_q/\partial q|^{-1}$ and the fermion-phonon coupling
$|f_{q,\ell}|^2$.

We first determine the mobility $\mu$ of the fermions in the Ohmic regime,
which we define by the relation $\bar{v}_d = \mu\delta$ in the limit
$\delta\rightarrow 0$, where $\delta = \Delta/4J_b$ is approximately the
ratio of the lattice tilt and the phonon bandwidth for the small values
of~$U_bn_b/J_b$ we consider here. We find that $\bar{v}_d/\delta \propto
\sum_{\ell=1}^\infty \ell^2\,\jj^2_\ell\left(\ell\,\zeta\right)$ for
$\delta\rightarrow 0$, with $\zeta = 2aJ_f/\hbar c_s$ as previously defined.
This sum is a Kapteyn series of the second kind, for which a closed form is
known~\cite{Watson-2008}, yielding for the mobility
\begin{equation}
  \mu = \frac{a U_{bf}^2}{8\sqrt{2}\,U_b\hbar}\left(\frac{J_b}{U_b n_b}\right)^{1/2}
  \frac{\zeta^2(\zeta^2+4)}{(1-\zeta^2)^{7/2}}\,.
\end{equation}
The mobility diverges in the limit $\zeta\rightarrow 1$, however, it is well
defined in the regime $\zeta\ll 1$. For the practically important case,
where the potential depth for the fermions and hence the hopping parameter
$J_f$ is varied, the mobility scales as $\mu\sim J_f^2$ in the limit
$J_f\rightarrow 0$.

\begin{figure}[t]
\begin{center}
  \includegraphics[width = 185pt]{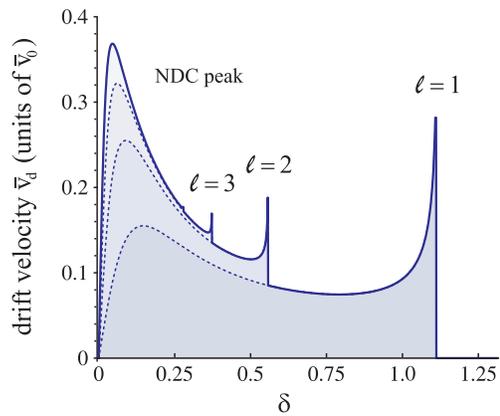}
  \caption{(Color online) The total drift velocity $\bar{v}_d$ (full line) as a function of the lattice tilt~$\delta=\Delta/4J_b$.
  The contributions from different jump distances~$\ell$ (dashed lines) result in a distinct NDC peak
  and a series of phonon resonances at the positions~$\delta\approx 1/\ell$.
  The system parameters are $U_b n_b/J_b=0.5$, $J_f/J_b = 0.4$ yielding for the phonon bandwidth $1.12\times 4J_b$
  and $\zeta = 0.5$. The drift velocity is given in units of $\bar{v}_0 = a n_b U^2_{bf}/4J_b\hbar$.}\label{current}
\end{center}
\end{figure}

The dependence of the drift velocity $\bar{v}_d$ given in Eq.~(\ref{vdres})
on the full range of accessible lattice tilts $\delta$ is shown in
Fig.~\ref{current}. We see that the atomic current makes a sharp transition
from Ohmic conductance to negative differential conductance. The main reason
is that the width of the Wannier-Stark states, proportional to $J_f/\Delta$,
and hence the overlap between states at different sites decreases as the
tilt~$\Delta$ is increased. This reduced overlap results in a low drift
velocity $\bar{v}_d$---in analogy to NDC in semiconductor
superlattices~\cite{Dohler-SSC-1975,Leo-2003}---and is determined by the
matrix elements $f_{q,\ell}$. The drift velocity $\bar{v}_d$ depends on the
tilt as $(J_f/\Delta)^{2\ell}$ for each jump distance~$\ell$ in the limit of
small fermionic bandwidth $J_f/\Delta\ll1$, which can be readily achieved in
cold atom systems. As can be seen in Fig.~\ref{current} the NDC is
particularly pronounced due to the superposed contributions from different
jump distances $\ell$ to the total current. These findings are in accordance
with predictions for a free homogeneous
superfluid~\cite{Ponomarev-PRL-2006,Bruderer-NJP-2008,Kolovsky-PRA-2008},
however, the result for $\bar{v}_d$ in Eq.~(\ref{vdres}) describes
additional features stemming from the influence of the phonon density of
states.

Specifically, the drift velocity exhibits anomalies in the form of sharp
peaks, which correspond to the anticipated electron-phonon resonances in a
solid-state system~\cite{Bryxin-SSC-1972,Bryksin-JPCM-1997}. As shown in
Fig.~\ref{phonon}(b), the phonon density of states
$|\partial\hbar\omega_q/\partial q|^{-1}$ is approximately constant for
small momenta, but exhibits a Van Hove singularity~\cite{VanHove-PR-1953} at
the edge of the first Brillouin zone. Therefore phonon resonances arise as
the momentum of the emitted phonons approaches the edge of the phonon band.
It can be found from Eq.~(\ref{vdres}) that the resonance condition is given
by $\ell\delta = \sqrt{1+U_bn_b/2J_b}$ with $\ell =
1,2,\ldots,\ell_{\mathrm{max}}$ and thus the current displays a peak at
$\delta\approx 1/\ell$ corresponding to each admissible jump
distance~$\ell$. Since the fermion-phonon interaction provides the only
relaxation process in the system these anomalies are directly reflected in
the tilt dependence of the current. Further, the current vanishes as the
lattice tilt exceeds the phonon bandwidth because there are no phonon states
available in order to dissipate energy into the superfluid.

Let us briefly discuss the effect of an additional (shallow) harmonic
trapping potential of the bosons on the phonon resonances. If the system is
aligned along the $x$-axis the potential takes the form $V(x) =
m_b\omega_x^2x^2/2$ with the trap frequency $\omega_x$. In the
experimentally relevant case, where the harmonic oscillator length $l_x =
\sqrt{\hbar/m_b\omega_x}$ satisfies the condition $l_x/N\gg n_ba$, the local
density approximation (LDA) is
applicable~\cite{Albus-PRA-2003,Kramer-EPJD-2003}. The LDA consists of
replacing the bosonic density $n_b$ by the position-dependent density
$n_b(x)$, determined by $V(x)$, with the other parameters of the model
unmodified. Consequently, the average drift velocity in Eq.~(\ref{vdres})
has to be replaced by $\bar{v}_d^{\mathrm{LDA}} = \int\dd
x\,p(x)\bar{v}_d[n_b(x)]$, where $p(x)$ denotes the normalized spatial
distribution of the fermions. This averaging might cause some broadening of
the phonon resonances. However, the position of the resonances $\delta
=\ell^{-1}\sqrt{1+U_bn_b/2J_b}$ and more generally $\bar{v}_d$ (except for
the prefactor $\bar{v}_0$) depend on $n_b$ only through the expression $U_b
n_b/J_b \ll 1$. Thus the trap-induced broadening can be made arbitrarily
small by reducing the boson-boson interaction energy $U_b$, which sets the
smallest energy scale close to resonance. This is partly explained by the
fact that the resonances involve only phonons from the upper edge of the
phonon band with wavelengths comparable to the lattice spacing $a$.


\section{Numerical simulation}

The theoretical results derived in the previous section are, strictly
speaking, valid for stationary atomic currents in a homogeneous system. We
now show based on numerical simulations that the predicted negative
differential conductance and the phonon resonances are observable in a
system of finite size under the conditions of a realistic measuring
procedure. In order to measure the atomic current in an experimental setup
we envisage a procedure consisting of the following three steps:
\begin{enumerate}[(a)]

  \item Initially, both the bosons and fermions are prepared in a
      horizontal optical lattice and the total system is in
      equilibrium~\footnote{ Our numerical simulations show that the
      total equilibration of the system is not an essential requirement
      for the procedure to work. We obtain similar results if the
      fermions are immersed into the superfluid in a nonadiabatic way.}.
      The fermions are each localized in separate sites sparsely
      distributed through the fermionic lattice of sufficient depth so
      that $P_i P_j \approx 0$ holds. In this configuration the fermions
      are automatically cooled by the surrounding
      superfluid~\cite{Daley-PRA-2004}, which is only slightly
      distorted~\cite{Bruderer-NJP-2008}.

  \item Subsequently, the lattice of the fermions is tilted for a fixed
      evolution time~$\tau$ of the order of $\hbar/J_f$ to let them
      evolve. To obtain a detectable displacement of the fermions the
      fermionic hopping $J_f$ may have to be increased by reducing the
      depth of the lattice.

  \item Finally, the spatial distribution of the fermions is detected,
      e.g.,~by \emph{in situ} single-atom resolved
      imaging~\cite{Bakr-NAT-2009,Sherson-AX-2010}. From the difference
      between the initial and final distributions it should be possible
      to extract a reliable estimate for the atomic current.
      Alternatively, the momentum distribution and hence the drift
      velocity of the fermions may be determined directly by a
      time-of-flight measurement.

\end{enumerate}

\begin{figure}[t]
\begin{center}
  \includegraphics[width = 185pt]{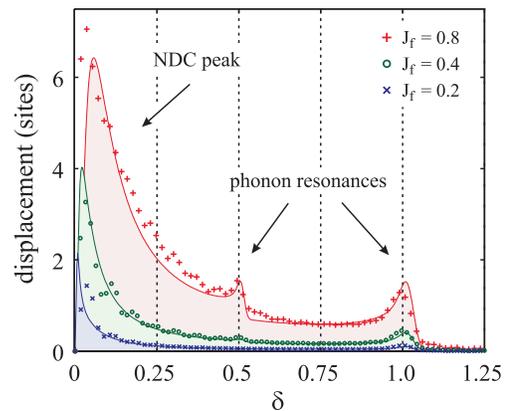}
  \caption{(Color online) The displacement of the fermion as a function of the lattice tilt~$\delta$
  determined numerically (symbols $+$, $\circ$, $\times$) and corresponding analytical fits (solid lines).
  The dominant NDC peak and the $\ell = 1$ phonon resonance at the boundary of the phonon
  band can be clearly recognized. The ratio $V$ between the height of the $\ell=1$ resonance and the
  NDC peak is $V=\{0.09, 0.14, 0.18\}$ for increasing $J_f$. The $\ell = 2$ resonance is visible for sufficiently
  large fermionic hopping~$J_f$. The system parameters are $J_f = \{0.2,0.4,0.8\}$,
  $U_b = 0.1$, $U_{bf} = 0.5$ and the fitting parameters are
  $\varepsilon = 0.1$, $\tilde{J}_f = 0.7J_f$, $\ell_{\mathrm{max}}=2$,
  with energies in units of $J_b$.}\label{fit}
\end{center}
\end{figure}

To demonstrate the feasibility of this procedure we have simulated the fully
coherent dynamics at zero temperature of both the bosons and fermions based
on the complete Bose-Fermi Hubbard model given in Eq.~(\ref{ham}). To this
end we have used the time-evolving block decimation (TEBD)
algorithm~\cite{Vidal-PRL-2003}, which is essentially an extension to the
well-established density matrix renormalization group (DMRG)
method~\cite{White-PRL-1992,Schollwock-RMP-2005}. The TEBD algorithm permits
the near-exact dynamical simulation of quantum many-body systems far from
equilibrium, which is crucial for our purposes. The experimental procedure
was simulated for a system consisting of $101$ lattice sites with a bosonic
filling factor $n_b = 1$ and a single fermion initially located at the
center of the lattice. We used box boundary conditions and the evolution
time was limited to $\tau = 24\hbar/J_b$ to minimize finite-size effects,
primarily reflections of phonons from the boundaries.

Figure~\ref{fit} shows the displacement of the fermion as a function of the
lattice tilt $\delta$ for a set of realistic experimental parameters, in
particular, for three different values of the fermion hopping $J_f$. The
main features predicted by our theoretical analysis can be clearly
recognized, namely the dominant NCD-peak, the phonon resonance at
$\delta\approx 1$ corresponding to the jump distance $\ell = 1$ and the
suppression of the current for $\delta\gg1$. In addition, for sufficiently
large values of $J_f$ the phonon resonance at $\delta\approx 1/2$
corresponding to the jump distance $\ell = 2$ is visible. For lower values
of $J_f$ the phonon resonance for $\ell = 2$ may still lead to a
characteristic drop in the current once the lattice tilt exceeds half the
phonon bandwidth. The resonances for higher values of the jump distance~$\ell$
seem to be masked mainly by broadening effects for the parameter regimes
tested.

The ratio between the height of the $\ell=1$ phonon resonance and the NDC
peak, i.e.,~their relative visibility $V$, depends nontrivially on the
fermion hopping~$J_f$. The hopping $J_f$ enters the expression for
$\bar{v}_d$ in Eq.~(\ref{vdres}) through the matrix elements $f_{q,\ell}$ in
the form of $\jj_\ell(s\ell J_f/J_b)$, where the parameter $s$ depends on
the lattice tilt. Close to resonance we have $s\approx1$, whereas for small
tilts $\ell\Delta\ll U_bn_b$ we obtain $s\approx\sqrt{2J_b/U_bn_b}$ and
hence $s\gg 1$ in the superfluid regime. Accordingly, the height of the
$\ell=1$ phonon resonance ($s\approx 1$) varies only slowly with $J_f$ as
opposed to the height of the NDC peak ($s\gg 1$), which is characterized by
the oscillatory nature of the Bessel functions. Thus a careful choice of
$J_f$ allows us to optimize the visibility $V$, i.e.,~to minimize the height
of the NDC peak, by tuning $s\ell J_f/J_b$ close to a zero of the relevant
Bessel functions. For the increasing values of $J_f$ considered in our
simulation the first zero of the Bessel function $\jj_2(z)$ is approached,
which explains the improved visibility $V$ for higher $J_f$, shown in
Fig.~\ref{fit}.

The general broadening of the phonon resonances may be caused by finite-size
effects and multiphonon processes. The finite-size effects include
reflections of phonons from the boundary of the system that in turn affect
the motion of the fermion. Multi-phonon processes, which have been neglected
in our theoretical analysis, also cause some broadening of the single-phonon
resonances. In particular the continuous drop in the current to zero can be
explained by the emission of several low-energy phonons during a jump
process, which is allowed even if the lattice tilt exceeds the phonon
bandwidth.

In order to compare the numerical and the theoretical results, at least
qualitatively, we adapt our expression for the drift velocity $\bar{v}_d$ to
a system of finite size. In addition, we introduce the observed broadening
effects characterized by the energy $\varepsilon$ into the theory.
Explicitly, we calculate the average drift velocity by using the expression
$\bar{v}_d^\varepsilon =
\sum_{\ell>0}^{\ell_{\mathrm{max}}}(a\ell)\,W_\ell^\varepsilon$ with the
rates $W_\ell^\varepsilon
=(2\pi/\sqrt{\pi}\varepsilon\hbar)\sum_q|f_{q,\ell}|^2\,\exp[-(\hbar\omega_q-\ell\Delta)^2/\varepsilon^2]$.
The rates reduce to the previous expression $W_\ell^0$ in absence of
broadening, i.e.,~$\lim_{\varepsilon\rightarrow 0}W_\ell^\varepsilon =
W_\ell^0$. The average displacement of the fermion (in units of lattice
sites) after the evolution time $\tau$ is then approximately given by
$\bar{x}_a \approx \bar{v}_d^\varepsilon\,\tau/a$.

We fit this theoretical model to the numerical results with the
broadening~$\varepsilon$, the maximal jump distance~$\ell_{\mathrm{max}}$,
and the \emph{effective} fermionic hopping $\tilde{J}_f$ as free parameters.
The first two parameters allow us to extract a quantitative value for the
broadening and to determine the dominant hopping process in the experiment.
The fermionic hopping needs adjustment because the bare hopping $J_f$ is
renormalized by coherent phonon processes, which lead to a reduced hopping
$\tilde{J}_f < J_f$ as discussed in
Refs.~\cite{Bruderer-PRA-2007,Bruderer-NJP-2008}. This is in direct analogy
to the increased effective mass of polarons due to the drag of the phonon
cloud~\cite{Mahan-2000}.

\begin{figure}[t]
\begin{center}
  \includegraphics[width = 235pt]{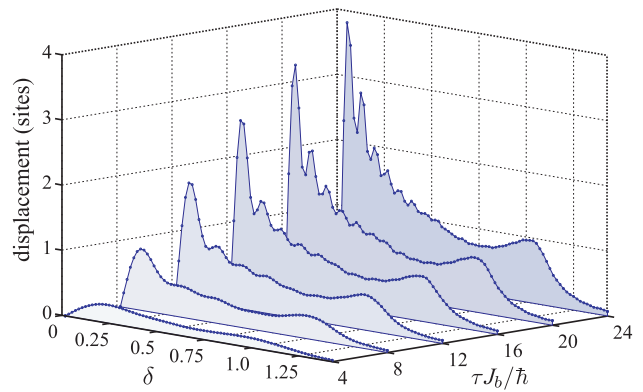}
  \caption{(Color online) The displacement of the fermion as a function of the lattice tilt~$\delta$
  for different evolution times $\tau$ determined numerically (connected data points).
  The NDC peak and the $\ell = 1$ phonon resonance are clearly visible also in
  presence of strong boson-fermion and boson-boson interactions. The additional peaks
  between the NDC peak and the phonon resonance are due to finite-time effects. The
  displacement increases approximately linearly with the evolution time~$\tau$.
  The system parameters are $J_f = 0.5$, $U_b = U_{bf} = 1$,
  with energies in units of~$J_b$.}\label{waterfall}
\end{center}
\end{figure}

As can be seen in Fig.~\ref{fit} the fit describes the numerical results
very accurately with moderate broadening~$\varepsilon$ and a minor reduction
of the fermionic hopping~$J_f$. Further, we find that $\ell_{\mathrm{max}} =
2$ and thus the dominant transport processes are nearest- and
next-nearest-neighbor hopping; this is consistent with the absence of
higher-order phonon resonances noted earlier. The high level of agreement
between our theoretical model and the numerical results is partly explained
by two observations: First, the numerical results show that the fermion
reaches a constant drift velocity on a time scale much shorter than the
evolution time~$\tau$, thus transient effects due to the sudden tilting of
the lattice are negligible. Second, the drift velocity remains approximately
constant over the entire evolution time, as illustrated in
Fig.~\ref{waterfall}, resulting in a stationary atomic current.

Since our numerics simulates the full dynamics of the system we can monitor
the average position of the fermion at different evolution times $\tau$ and
extend the range of parameters considered so far. In particular, we are
interested in the atomic current in presence of strong boson-fermion and
boson-boson interactions, where our theory, assuming a superfluid phase, is
no longer applicable. Figure~\ref{waterfall} shows the displacement of the
fermion as a function of the lattice tilt~$\delta$ for the interaction
strengths $U_b = U_{bf} = J_b$ and for different evolution times~$\tau$. We
see that the NDC peak and the $\ell = 1$ phonon resonance are not noticeably
affected by the presence of strong interactions, thus they both seem to be
robust features of the atomic current. We note that the additional peaks in
the displacement-tilt dependence (located between the NDC peak and the $\ell
= 1$ phonon resonance) are due to finite-time effects, which is revealed by
a more detailed analysis of the time dependence of the numerical results.
Furthermore, we observe an approximately linear increase of the displacement
of the fermion with the evolution time---a finding also confirmed for weaker
interactions.


\section{Conclusions}

In our analytical and numerical investigation of phonon-assisted atomic
currents along a tilted potential we have shown that Bose-Fermi mixtures in
optical lattices lend themselves naturally to investigate nonequilibrium
transport phenomena present in solid-state systems. In more detail, we have
formulated an effective model for the Bose-Fermi mixture describing the
bosons and fermions in terms of Bogoliubov phonons and Wannier-Stark states,
respectively, with a generic fermion-phonon interaction of an identical type
to the one encoutered in solids.

We have studied the dependence of the atomic current on the lattice tilt
from first principles and found that our model accommodates negative
differential conductance and phonon resonances. To demonstrate that these
features are observable by using ultracold atoms in the context of a
finite-size system and a realistic measuring procedure we have calculated
the atomic current numerically by using the TEBD algorithm including the
full dynamics of both the bosons and the fermions. Our numerical results
show that the phonon resonance at the boundary of the phonon band is a
robust phenomenon that occurs over a wide range of system parameters despite
broadening, which might be increased by finite temperature
effects~\cite{Bruderer-NJP-2008,Kolovsky-PRA-2008} not directly taken into
account in our model. Finally, we note that in an obvious extension of this
work we will investigate the effect of the transition of the bosons from the
superfluid to the Mott insulator regime on the atomic current through the
system.


\acknowledgements

M.B. thanks the Swiss National Science Foundation for the support through
the project PBSKP2/130366. M.B., W.B. and A.P. acknowledge financial support
from the German Research Foundation (DFG) through SFB 767. S.R.C. and D.J.
thank the National Research Foundation and the Ministry of Education of
Singapore for support. D.J. acknowledges support from the ESF program
EuroQUAM (EPSRC Grant No. EP/E041612/1).

\renewcommand{\baselinestretch}{1.05}
\bibliography{resonance}

\end{document}